\documentclass[english,preprint,showpacs,preprintnumbers,amsmath,amssymb]{revtex4}
\usepackage[T1]{fontenc}
\usepackage[latin1]{inputenc}
\usepackage{graphicx}
\usepackage{amssymb}

\makeatletter

\usepackage{amsfonts}

\usepackage{epsfig}

\usepackage{dcolumn}

\usepackage{bm}

\usepackage{babel}
\makeatother
\begin{document}

\title{Custodial Symmetry and Extensions of the Standard Model}
\author{J. C. Montero}
\email{montero@ift.unesp.br}
\affiliation{Instituto de F\'{\i}sica Te\'{o}rica, Universidade Estadual Paulista\\
 Rua Pamplona 145, 01405-000, S\~{a}o Paulo, SP --- Brazil. }
\author{V. Pleitez}
\email{vicente@ift.unesp.br}
\affiliation{Instituto de F\'{\i}sica Te\'{o}rica, Universidade Estadual Paulista\\
 Rua Pamplona 145, 01405-000, S\~{a}o Paulo, SP --- Brazil. }

\date{\today}

\begin{abstract}
We show that the extension of the approximate custodial
$SU(2)_{L+R}$ global symmetry to all the Yukawa interactions of the
standard model Lagrangian implies the introduction of sterile
right-handed neutrinos and the seesaw mechanism in this sector. In
this framework, the observed quark and lepton masses may be
interpreted as an effect of physics beyond the standard model. The
mechanism used for breaking this symmetry in the Yukawa sector could
be different from the one at work in the vector boson sector. We
give three model independent examples of these mechanisms.
\end{abstract}


\maketitle

\section{Introduction}

\label{sec:intro}

One of the most important problems in elementary particle physics is
concerned about the origin of the fermion masses. Part of the
problem, that we will consider here, can be formulated as follows:
why do weak isospin partners have different masses? It is well known
that, in the context of the standard model (SM), the weak isospin
partners have different masses because the left- and right-handed
components of the fields belong to different re\-pre\-sen\-ta\-tion
of the $SU(2)_{L}\otimes U(1)_{Y}$, and they are not related by a
symmetry~\cite{volkas}. On the other hand, we may wonder about the
\textit{natural} values for the fermions masses in the context of
the SM. For the light quarks ($u,d$ and $s$) their masses are well
below the energy scale, $\Lambda_\chi\sim\Lambda_{QCD}$, of the
chiral $SU(3)_L\otimes SU(3)_R$ symmetry breaking and the
spontaneous breakdown of this symmetry dominates the dynamics of
these flavors. We can say that their masses are natural in the sense
that vanishing them will augment the symmetry of the model. For
heavy quarks, whose masses are larger than $\Lambda_\chi$, the
explicit breaking of the chiral symmetry dominates the
dynamics~\cite{manohar}. On the other hand, in the limit of exact
global $SU(2)_{L+R}$ isospin symmetry we have that $m_{u}=m_{d}$.
The fact that $m_u\approx m_d$ means that the isospin symmetry is
not so badly broken. The case of $SU(3)_{L+R}$ flavor symmetry is
not as good realized as the $SU(2)_{L+R}$ but is better than the
other higher flavor global chiral symmetries like $SU(4)_{L+R}$ etc.
Moreover, from the point of view of the electroweak spontaneous
symmetry breaking, we would expect the quark masses to be of the
order of magnitude of the symmetry breaking parameter
$v=(\sqrt{2}/2G_F)^{1/2}\sim246~$~GeV. Only the quark top satisfies
this requirement since $m_t/v\approx 1/\sqrt2$.

In this work we will assume that the natural values for all quarks
and leptons are those in which weak isospin partners have equal
masses. This assumption is not in conflict with the discussion about
the light and heavy quarks given above; it is in fact complementary.
Hence, we have $M_u= M_d$, $M_l= M^{Dirac}_{\nu_l}$, where
$M_u(M_d)$ denotes the mass matrix for the 2/3(-1/3) charged quarks
and $M_l(M_{\nu_l})$ denotes the mass matrix for the charged
leptons(neutrinos). Notice that at this point the neutrino masses
are of Dirac type. If this where the case, the observed mass spectra
$m_u\not= m_d$, $m_c\not= m_s$, etc., must be considered already an
effect of physics beyond the SM. In the lepton sector whatever is
the new physics it must include the seesaw mechanism for generating
small masses for the active neutrinos.

Independently of the problem above, it was noted in the late
seventies that by ignoring the electromagnetism in the electroweak
sector of the SM, $g^\prime=0\;(\sin\theta_W=0)$,  the $W$ boson
fields $W^{\pm},W^{3}\equiv Z$ transform as triplet under a global
$SU(2)_{L+R}$ symmetry and, as a consequence, the masses of the
three vector bosons are equal: $M_{Z}=M_{W^{\pm}}$~\cite{veltman}.
The $\rho$-parameter defined as $\rho\equiv
M^2_W/M^2_Z\cos^2\theta_W$  has in this limit ($\cos\theta_W=1$) the
value $\rho=1$. The inclusion of electromagnetism through the
$U(1)_Y$ factor i.~e., $g^\prime\not=0$ (or $\sin\theta_W\not=0$),
breaks the global symmetry and the $W^{3}$--$B$ mixing introduces
the zero order (in the sense that it does not depend neither on $g$,
nor on $g^\prime$) correction factor $\cos^2\theta_{W}\not=1$ to the
ratio of the vector boson masses but still we have $\rho=1$ at the
tree level. When radiative corrections are considered
$\rho\approx1$. That global symmetry can be in some cases identified
with the $SU(2)$ isospin, and since it protects the value of $\rho$
from radiative corrections it is called custodial symmetry. This
sort of symmetry has been also explored in other
contexts~\cite{custodial,new}.

Let us review how the $SU(2)$ custodial symmetry has been
implemented in the scalar sector of the SM. This model has only one
Higgs scalar doublet $H=(h^{+}\, h^{0})^{T}$ transforming as
$(\textbf{2},+1)$ under $SU(2)_{L}\otimes U(1)_{Y}$. By introducing
the 2-doublet~\cite{will}
\begin{equation}
\Phi=\frac{1}{\sqrt{2}}(H\,\tilde{H}),\label{2doublet}\end{equation}
where $\tilde{H}=\epsilon H^{*}$, we can write the usual
$SU(2)_{L}\otimes U(1)_{Y}$
invariant Lagrangian of the Higgs scalar sector in terms of the 2-doublet
\begin{equation}
\mathcal{L}=\textrm{Tr}[(D_{\mu}\Phi)^{\dagger}D^{\mu}\Phi]-
V(\Phi^{\dagger}\Phi), \label{higgs}
\end{equation}
where $V(\Phi^{\dagger}\Phi)=-\mu^{2}\textrm{Tr}\Phi^{\dagger}\Phi+
\lambda(\textrm{Tr}\Phi^{\dagger}\Phi)^{2}$
and the covariant derivative becomes
\begin{equation}
D_{\mu}\Phi=\left(\partial_{\mu}\Phi+i\frac{g}{2}\vec{W}_{\mu}\cdot
\vec{\sigma}\,\Phi+i\frac{g^{\prime}}{2}B_{\mu}\Phi\sigma_{3}\right).
\label{cderivada}
\end{equation}

We see that the Lagrangian in Eq.~(\ref{higgs}) is invariant under
the local and global $SU(2)_{L}\otimes U(1)_{Y}$ symmetry: $\Phi\to
L\Phi$ and $\Phi\to\Phi \exp(-i\sigma_3/2)$. However in the limit
when $g^{\prime}=0$ we have a larger right-handed symmetry:
$\Phi\to\Phi R^{\dagger}$ where $R$ is a global $SU(2)_{R}$
transformation. In this case we have that
$\Phi\sim(\textbf{2},\textbf{2})$ under global $SU(2)_{L}\otimes
SU(2)_{R}$. The Lagrangian in Eq.~(\ref{higgs}) has an approximate
(valid when $g^{\prime}=0$) $SU(2)_{L}\otimes SU(2)_{R}$ global
symmetry which is broken to $SU(2)_{L+R}$ when the neutral component
of $H$ gains a vacuum expectation value,
$\langle\Phi\rangle=\textrm{diag}(v, v)/\sqrt2$. Now $L=R$ and
$\langle\Phi\rangle\to L\langle\Phi\rangle L^\dagger$ under
$SU(2)_{L+R}$. Next, $SU(2)_{L+R}$ is broken when $g^\prime\not=0$.

In the next section, motivated by the idea that it may be natural
that, in the context of pure SM, the weak isospin partners have
equal masses, we will extend the custodial symmetry to all the
Yukawa sectors of the SM. Within the minimal SM representation
content, i.~e., no sterile neutrinos at all, that symmetry can be
implemented with the known fermions only in the quark sector. Quarks
are strong interacting particles so that it makes sense to consider
the turning off of the electromagnetic interactions. Moreover, by
adding right-handed neutrinos we are able to extend this approximate
custodial symmetry also to the Yukawa lepton sector. In this sector
we may consider that the equivalent to the ``strong interactions'',
with respect to which we can turn off the electromagnetic
interactions, are the large Majorana masses of the right-handed
neutrinos. The latter ones are sterile with respect to the SM
interactions, and trigger the type--I seesaw
mechanism~\cite{seesaw}. Hence, this mechanism is mandatory in the
scheme.

\section{The custodial symmetry in the Yukawa sector}
\label{sec:custodialsm}

The condition under which isospin partners like proton and neutron
have the same mass was put forward by Weinberg in the early
seventies in the context of an $SU(2)_L\otimes SU(2)_R\otimes U(1)$
model~\cite{sw72}. This is straightforward extended for $u$ and $d$
quarks. The generalization of this symmetry to the other weak
isospin doublets implies that this equality is valid for any 2/3 and
-1/3 charged quarks and similarly in the lepton sector. As we said
in the previous section the extension of the custodial symmetry to
the Yukawa sector, as it was suggested recently in another
context~\cite{newpprd}, implies that the weak-isospin partners have
the same mass. On the other hand, in order to be able to implement
this symmetry in the Yukawa leptonic sector we ought to introduce
right-handed neutrinos. If we consider that the weak isospin
partners have equal masses the observed fermion masses should be
already seen as an effect of new interactions that break explicitly,
or not, the custodial symmetry. We give some model independent
examples of possible new physics breaking the custodial symmetry and
producing the observed fermion masses. They are: i) the existence of
extra fermions transforming as singlets under local $SU(2)_{L}$, ii)
several scalar doublets, and/or, iii) new neutral vector bosons with
non-universal couplings to fermions.

As usual we define the minimal representation content in the
standard model \textit{plus} right-handed neutrinos: the doublets
$L_{aL}=~(\nu_{aL}\, l_{aL})^{T}\sim(\textbf{2},-1)$,
$Q_{iL}=(u_{iL}\, d_{iL})^{T}\sim(\textbf{2},1/3)$, and the singlets
$l_{aR}\sim(\textbf{1},-2)$, $u_{iR}\sim(\textbf{1},4/3)$,
$d_{iR}\sim(\textbf{1},-2/3)$, and $\nu_{aR}\sim(\textbf{1},0)$,
where $a=e,\mu,\tau$ and $i=1,2,3$. Notice that it is only because
we have included right-handed neutrinos that we are able to define
in the lepton sector the 2-doublets, $(\textbf{2},\textbf{2})$ under
$SU(2)_{L}\otimes SU(2)_{R}$,
$\textrm{L}_{ab}=~(\overline{L}_{a}l_{bR}\;\overline{L}_{a}\nu_{bR})$.
In the quark sector however we already have all the fields needed to
write the 2-doublets $\textrm{Q}_{ij}=(\overline{Q}_{iL}d_{jR}\;
\overline{Q}_{iL}u_{jR})$. Both 2-doublets $\textrm{L}_{ab}$ and
$\textrm{Q}_{ij}$ transform under $SU(2)_{L+R}$ as
$\langle\Phi\rangle$ in the last section. Thus, we can write the
Yukawa interactions which are invariant under this global symmetry
as follows:
\begin{equation}
-\mathcal{L}_{Y}^{\prime}=h_{ab}\textrm{Tr}(\textrm{L}_{ab}\Phi)+
g_{ij}\textrm{Tr}(\textrm{Q}_{ij}\Phi)+H.c.
 \label{yuka1}
\end{equation}
We note from the Yukawa interactions in Eq.~(\ref{yuka1}) that if
the custodial sy\-mme\-try is not broken the mass of the charged
lepton in each generation is equal to the mass of the respective
neutrino. The same happens in the quark sector: the $u$-type quarks
have the same mass of the $d$-type quarks, for each generation.
Explicitly we have
$(M_{l})_{ab}=(M^{Dirac}_{\nu_l})_{ab}=h_{ab}v/\sqrt{2}$; and
$(M_{u})_{ij}=(M_{d})_{ij}=g_{ij}v/\sqrt{2}$. After diagonalizing
these matrices we obtain the unrealistic mass relations
$\hat{M}_{l}=\hat{M}^{Dirac}_{\nu_l}=(v/\sqrt{2})\textrm{diag}(h_{e},
\,h_{\mu},\,h_{\tau})$; and
$\hat{M}_{u}=\hat{M}_{d}=(v/\sqrt{2})\textrm{diag}(g_{1},\,
g_{2},\,g_{3})$. It is interesting that in $SU(2)_L\otimes
SU(2)_R\otimes U(1)$ gauge models the custodial symmetry implies
relations between the masses of quarks and leptons of each
generation~\cite{volkas}.

Thus, we have to break the custodial symmetry in order to generate
the observed fermion masses. In the context of the minimal SM i.~e.,
no sterile neutrinos at all, this is not a problem because there is
no a custodial symmetry in the lepton sector since, as we said
before, there is no enough fields to define 2-doublets as the
$\textrm{L}_{ab}$ above. The custodial symmetry in the boson sector
is broken by turning on the electromagnetic interactions i.~e.,
$g^{\prime}\not=0$, and also by the Yukawa interactions, with or
without right-handed neutrinos, if we write them as
\begin{equation}
-\mathcal{L}_{Y}=\left(\overline{Q}_{iL}\Gamma_{ij}^{u}u_{jR}+
\overline{L}_{aL}\Gamma_{ab}^{\nu}\nu_{bR}\right)\tilde{H}+
\left(\overline{Q}_{iL}\Gamma_{ij}^{d}d_{jR}+\overline{L}_{aL}
\Gamma_{ab}^{l}l_{bR}\right)H+H.c.,
\label{yuka2}
\end{equation}
where $i,j=1,2,3$; $a,b=e,\mu,\tau$ and we have here, and below,
omitted summation symbols. If all Yukawa couplings $\Gamma$s in
Eq.~(\ref{yuka2}) are different, the generated Dirac masses in each
charged sector are different and arbitrary. The breaking mechanism
of the custodial symmetry in Eq.~(\ref{yuka2}) is the same as in the
boson sector: $g^{\prime}\not=0$.

However, instead of assuming the interactions in Eq.~(\ref{yuka2}),
we can look for different ways to obtain arbitrary masses and mixing
matrices, in all charged sectors, starting from Eq.~(\ref{yuka1}).
Then, we will look for mechanisms that can produce realistic masses
for all fermions in the standard model by breaking the custodial
symmetry in at least three model independent examples: \textit{i)} a
generalized seesaw mechanism~\cite{davidson,tau}; \textit{ii)}
multi-Higgs doublet-like SM extensions; and \textit{iii)} radiative
corrections mechanism. The latter case involves extra neutral vector
bosons with non-universal couplings to leptons. In all these cases
the seesaw mechanism is necessary in the neutrino sector.

\subsection{Generalized seesaw mechanism}
\label{subsec:gs}

One way to obtain a general mass matrix for quarks and leptons is to
add at least one extra $u$- and $d$-type quarks, $e^{-}$-type
charged leptons, and three sterile neutrinos $\nu_R$. These fields
transform under $SU(3)_{C}\otimes~SU(2)_{L} \otimes U(1)_{Y}$ as:
$U_{L,R}\sim(\textbf{3},\textbf{1},4/3)$,
$D_{L,R}~\sim~(\textbf{3},\textbf{1},-2/3)$,
$E_{L,R}\sim(\textbf{1}, \textbf{1},-2)$, and
$\nu_R\sim(\textbf{1},\textbf{1},0)$, respectively. In this way the
Yukawa interactions to be added to Eq.~(\ref{yuka1}), for breaking
explicitly the custodial symmetry, are
\begin{eqnarray}
-\mathcal{L}_{Y} & = & \overline{Q_{iL}}\,g_{_{iU}}U_{R}\tilde{H}+
\overline{Q_{iL}}\,g_{_{iD}}D_{R}H+\overline{U_{L}}\,M_{_{Ui}}u_{iR}+
\overline{D_{L}}\,M_{_{Di}}d_{iR}+M_{_{UU}}\overline{U_{L}}\,U_{R}\nonumber \\
& + &
M_{_{DD}}\overline{D_{L}}\,D_{R}+\overline{L_{aL}}\,h_{aE}E_{R}H+
M_{EE}\overline{E_{L}}\,E_{R}+\frac{1}{2}\,
\overline{(\nu_{aR})^{c}}\,(M_{R})_{ab}\nu_{bR}+H.c.,
\label{yukawasm2}
\end{eqnarray}
where $i,j=1,2,3$; $a,b=e,\mu,\,\tau$. The Yukawa coupling matrices
in the $u$-like quarks and charged leptons are, respectively,
\begin{eqnarray}
\Gamma_{u}=\left(\begin{array}{cccc}
g_{11} & g_{12} & g_{13} & g_{_{1U}}\\
g_{21} & g_{22} & g_{23} & g_{_{2U}}\\
g_{31} & g_{32} & g_{33} & g_{_{3U}}\\
\frac{\sqrt{2}}{v}M_{_{U1}} & \frac{\sqrt{2}}{v}M_{_{U2}} &
\frac{\sqrt{2}}{v}M_{_{U3}} & \frac{\sqrt{2}}{v}M_{_{UU}},
\end{array}\right)\!\!,\Gamma_{l}=\!\!\left(
\begin{array}{cccc}
h_{ee} & h_{e\mu} & h_{e\tau} & h_{_{eE}}\\
h_{\mu e} & h_{\mu\mu} & h_{\mu\tau} & h_{_{\mu E}}\\
h_{\tau e} & h_{\tau\mu} & h_{\tau\tau} & h_{_{\tau E}}\\
\frac{\sqrt{2}}{v}M_{_{Ee}} & \frac{\sqrt{2}}{v}M_{_{E\mu}} &
\frac{\sqrt{2}}{v}M_{_{E\tau}} &
\frac{\sqrt{2}}{v}M_{_{EE}}\end{array} \right)\!\!\!,\,
\label{seesawq}
\end{eqnarray}
and similarly, the $d$-like quark Yukawa matrix is obtained by
replacing in $\Gamma_{u}$, $g_{_{iU}}\to g_{_{iD}}$, $M_{_{Ui}}\to
M_{_{Di}}$ and $M_{_{UU}}\to M_{_{DD}}$. The related mass matrices
can implement a seesaw--like mechanism if $M_{_{UU}}\gg
g_{ij}v/\sqrt{2},\; g_{_{iU}}v/\sqrt{2},M_{_{Ui}}$, and similar
conditions in the other charged sectors. In fact, a particular case
of the Yukawa matrix for the charged leptons $\Gamma_{l}$, where the
sub-matrix $h_{ab}$ is antisymmetric, was worked out in
Ref.~\cite{tau} showing that it is possible to generate the observed
charged lepton masses, with a fine tuning which is not worst than
the usual one in the SM. In the neutrino sector we have the
$6\times6$ mass matrix
\begin{equation}
-\mathcal{M}_{\nu}=\frac{1}{2}\,\left(\begin{array}{cc}
0 & \frac{v}{\sqrt{2}}\, h\\
\frac{v}{\sqrt{2}}\, h^{T} & M_{R}\end{array}\right),
\label{nusmass}
\end{equation}
where $M_{R}$ is a general $3\times3$ matrix. With this scheme we
obtain a general mixing between left-handed neutrinos which allows
an arbitrary mixing matrix in the charged currents coupled to
$W^{\pm}$. We can verify also that the neutrino masses compatible
with neutrino oscillation data implies the existence of at least two
large Majorana mass scales. Assuming for simplicity that
$(v/\sqrt{2}\,)h\to(v/\sqrt{2}\,)\textrm{diag}(h_{e},
h_{\mu},h_{\tau})\approx\textrm{diag}(m_{e},m_{\mu},m_{\tau})$, and
$M_R=\textrm{diag}(M_1,M_2,M_3)$, we obtain $M_{1}\sim
M_{2}=10^{9}$~GeV and $M_{3}=10^{12}$~GeV.

In this case the mixing matrices in the quark and lepton sectors are
not exactly unitary and there is also flavor changing neutral
currents (FCNC) coupled to the $Z^{0}$ and the neutral scalars.
However, these sort of currents are naturally suppressed by the mass
of these extra fermions. For a general analysis see
Refs.~\cite{ll88}. On the other hand, the value of the $\rho$
parameter is still protected from radiative corrections since the
charged fermion singlets do not contribute to $\rho$ because they
couple to $Z^{0}$ only through vector currents. We have verified,
using the general expressions given by Veltman~\cite{veltman77},
that $\delta\rho|_{\textrm{singlets}}=0$, at least at the one loop
level. Anyway, this issue has to be considered in a particular model
in which the $SU(2)_L\otimes U(1)_Y$ symmetry and the extra fermions
may be imbedded~\cite{nusright,pp}.

\subsection{Multi-Higgs doublet-like extensions of the standard model}
\label{subsec:multi}

Another mechanism for breaking the custodial $SU(2)_{L+R}$ symmetry
is implemented by adding extra scalar doublets to the SM. Let us
suppose that we add four Higgs doublets $H_{i}=~(h_{i}^{+},\,
h_{i}^{0})^{T},\, i=1,3$ and $H_{j}=(h_{j}^{0},\,-h_{j}^{-})^{T},\,
j=2,4$, allowing to define the 2-doublets $\Phi_{12}=(H_{1},\,
H_{2})/\sqrt{2}$, and $\Phi_{34}=(H_{3},\, H_{4})/\sqrt{2}$. Other
($H_{i},\, H_{j})$ combinations may be avoided by imposing
appropriate discrete symmetries. In this case we have
\begin{equation}
-\mathcal{L}_{Y}^{\prime\prime}=h_{ab}\textrm{Tr}(\textrm{L}_{ab}
\Phi_{12})+h_{ab}^{\prime}\textrm{Tr}(\textrm{L}_{ab}\Phi_{34})+
g_{ij}\textrm{Tr}(\textrm{Q}_{ij}\Phi_{12})+g_{ij}^{\prime}\textrm{Tr}
(\textrm{Q}_{ij}\Phi_{34})+H.c.\label{yuka3}\end{equation}
 Denoting $\langle h_{i}^{0}\rangle=v_{i}/\sqrt{2}$ and considering
$v_{i}\not=v_{j}$, for $i\not=j$, we break the custodial symmetry,
and obtain the mass matrices \begin{equation}
(M_{l})_{ab}=\frac{1}{\sqrt2}\,\left[hv_{2}+
h^{\prime}v_{4}\right]_{ab},\quad(M^{Dirac}_{\nu_l})_{ab}=
\frac{1}{\sqrt2}\,\left[hv_{1}+ h^{\prime}v_{3}\right]_{ab},\,
\label{ufa3}
\end{equation}
\begin{equation}
(M_{d})_{ij}=\frac{1}{\sqrt2}\,\left[gv_{2}+
g^{\prime}v_{4}\right]_{ij},\quad(M_{u})_{ij}=
\frac{1}{\sqrt2}\,\left[gv_{1}+ g^{\prime}v_{3}\right]_{ij},
\label{ufa4}
\end{equation}
where we are assuming real vacuum expectation values for simplicity.

The point here is that because all the vacuum expectation values are
different from each other, the mass matrices $M_{l}$ and
$M^{Dirac}_{\nu_l}$ are numerically different so that they can be
diagonalized by quite different unitary matrices. This sort of
models have FCNC which implies fine tuning in some of the couplings
since the Higgs scalar can not be arbitrarily heavy. A more
economical model with natural FCNC suppression~\cite{gw77} is the
one in which one of the two 2-doublets, $\Phi_{12}$, couples only to
leptons, and the other, $\Phi_{34}$, couples only to quarks. Then
$M_l=hv_2/\sqrt2$, $M^{Dirac}_{\nu_l}=hv_1/\sqrt2$, $M_d=g^\prime
v_4/\sqrt2$, and $M_u=g^\prime v_3/\sqrt2$. In this case the Dirac
masses of the neutrinos are different from the charged lepton masses
and, if we assume that the Majorana mass matrix for the right-handed
neutrinos $M_R$ is already diagonal, there is no a realistic mixing
in the leptonic sector. However, if $M_R$ is not diagonal the active
neutrino mass matrix is
$M^\nu_{active}\approx-M^{Dirac}_{\nu_l}M^{-1}_R(M^{Dirac}_{\nu_l})^T$,
which is numerically different from $M^{Dirac}_{\nu_l}$, and
realistic mixing in the lepton sector may arise. But, if even in
this case the mixing is not the required one, we can still apply the
mechanism of radiative corrections, discussed in the next
subsection, in order to generate the appropriate mixing matrix in
the lepton sector and we will get
$\tilde{M}^\nu_{active}\approx-\tilde{M}^{Dirac}_{\nu_l}
M^{-1}_R(\tilde{M}^{Dirac}_{\nu_l})^T$ (see below).

\subsection{Radiative correction breakdown of the custodial symmetry}
\label{subsec:radiative}

Here we will assume again that the Yukawa interactions at the tree
level are those in Eq.~(\ref{yuka1}), i.~e.,
$(\Gamma_{l}^{\prime})_{ab}= (\Gamma_{\nu}^{\prime})_{ab}\equiv
h_{ab}$, $(\Gamma_{u}^{\prime})_{ij}=(\Gamma_{d}^{\prime})_{ij}
\equiv g_{ij}$. Next, let us assume that the correct quark masses
have already been obtained by some of the two mechanisms considered
above but lepton weak isospin partners are still mass degenerate.
Then, we have to break only the degeneracy between the charged
lepton and the Dirac neutrino masses in order to have an appropriate
mixing in this sector. The question is that, if the Majorana mass
matrix for right-handed neutrinos is assumed diagonal, the charged
lepton and neutrino mass matrices are diagonalized by the same
unitary matrices and, as discussed above, in this case there is no
mixing in the leptonic sector $U_{PMNS}=U^\dagger_{lL} U_{\nu L}$.
But in the present case since $M_l=M^{Dirac}_{\nu_l}$ then
$U_{lL}=U_{\nu L}$, and thus we may have no enough mixing among
active neutrinos. Thus, we have to break the mass degeneracy, or
induce more corrections to the lepton masses, in order to have an
appropriate mixing matrix. The small masses for the active neutrinos
will arise through the seesaw mechanism as in the other cases. In
this case the custodial symmetry breakdown of the SM can be
implemented by radiative corrections involving fields that are not
present in this model.

A realization of this mechanism implies the existence of extra
neutral vector bosons, say $Z^\prime$, with non-universal couplings
with right-handed leptons, which can induce a general finite mass
contribution for these particles. This sort of models already exist
in literature, see for instance Ref.~\cite{zprime}. The extra
neutral vector boson couples to active left-handed neutrino only
through the mixing of $Z$ with $Z^{\prime}$ which may arise at the
tree level or at higher orders. This is the case of models in which
left-handed and right-handed neutrino components transform in
different representation with respect to the symmetry that include
the $Z^{\prime}$. The most simple case is an extra $U(1)$ symmetry
under which only the right-handed neutrinos are active. The
calculability of the masses is used in the sense  of
Refs.~\cite{sw72,georgiglashow}. In this case there are diagrams,
like the one in Fig.~1, that generate finite mass contributions for
neutrinos and similarly for charged leptons. In the figure the
tadpole denotes $\langle h^0\rangle=v/\sqrt2$. We need here only
that $Z^{\prime}$ has non-universal couplings with right-handed
leptons, $f_{R}^{l},f_{R}^{\nu}$. Contributions to the matrix
$h_{ab}$, arising from Fig.~1, are proportional (up to logarithm
factors) to
$\textrm{g}_{aL}^{\nu}f_{bR}^{\nu}M^{Dirac}_{\nu_l}\Delta$, where
$\Delta~\equiv~(v^2/\sqrt{2})/(M^2_Z+M^2_{Z^{\prime}})$ and
$(M^{Dirac}_{\nu_l})_{ab}~=~h_{ab}v/\sqrt{2}$. On the other hand,
contributions to the mass matrix of the charged leptons are
proportional to $\textrm{g}_{aL}^{\, l}f_{bR}^{\, l}
M^{Dirac}_{\nu_l}\Delta$. If only right-handed neutrinos and the
Higgs scalar carry the charge related to $Z^\prime$, the diagram in
Fig.~1 is the only one in the language in which the gauge symmetry
breaking is introduced explicitly by tadpoles~\cite{georgiglashow}.
In this way we can generate arbitrary contributions for the masses
of both charged lepton sectors.

Then, in the lepton sector the mass matrices, after taken into
account the radiative corrections, become
\begin{eqnarray}
&&(\tilde{M}_{l})_{ab}\approx
(M^{Dirac}_{\nu_l})_{ab}+(2s_{W}^{2}-1)
(M^{Dirac}_{\nu_l})_{ac}(f_{R}^{l})_{cb}\, \Delta,\nonumber \\ &&
 (\tilde{M}^{Dirac}_{\nu_l})_{ab}\approx
(M^{Dirac}_{\nu_l})_{ab}+(M^{Dirac}_{\nu_l})_{ac}(f_{R}^{\nu})_{cb}\,
\Delta, \label{ufa2}
\end{eqnarray}
where $f_{L}^{l}$ and $f_{R}^{\nu}$ are general, model dependent,
non-diagonal matrices. Since the mixing of $Z$ with heavier
$Z^\prime,...$ has to be small, say
$\Delta~\lesssim~O(10^{-3})$~\cite{valencia}, $\tilde{M}_{l}$ and
$\tilde{M}^{Dirac}_{\nu_l}$ are still of the same order of magnitude
but numerically they can differ enough, depending of the values of
$f_{R}^{l},f_{R}^{\nu}$, to be diagonalized by different unitary
matrices in such a way that the mixing matrix in the charged
currents coupled to the boson $W$ does appear. The fact that
$\tilde{M}_{l}\approx \tilde{M}^{Dirac}_{\nu_l}$ is not a problem
because now we can implement the seesaw mechanism and the total
neutrino mass matrix has the form as in Eq.~(\ref{nusmass}). This
mechanism is useful for generating a realistic mixing in the lepton
sector when the 2 Higgs doublets extension of the SM with no FCNC is
considered as in the previous subsection. In this case, although the
masses of neutrinos and charged leptons are numerically different,
both mass matrices are diagonalized by the same unitary matrices
since $M_l=M^{Dirac}_{\nu_l}(v_2/v_1)$. When radiative corrections
are considered, the latter relation is not valid anymore.

In general, in models with extra neutral vector bosons with
non-universal interactions there are also FCNC that impose
constraints on the parameters of the models, see for instance
Refs.~\cite{valencia,cheung}.

\begin{figure}[ptb]
\begin{centering}
\includegraphics[width=10cm,height=4.5cm]{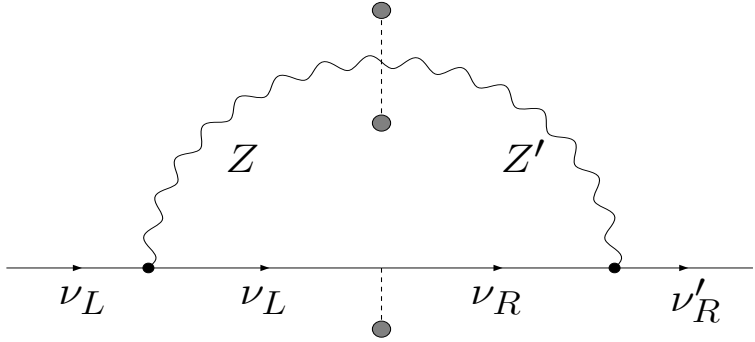}
\par
\end{centering}
\caption{\label{fig1} Possible finite contributions for the neutrino
Dirac masses. Similar diagrams do exist for charged leptons.}
\end{figure}

\section{Conclusions}
\label{sec:con}

We have seen that if we want to extend the custodial symmetry to the
Yukawa sector in the standard model it is mandatory to introduce
right-handed neutrinos and hence the seesaw mechanism. We have
considered three different model independent ways, involving new
physics, to break that symmetry, generating the appropriate masses
in the fermion sector. Our analysis was model independent but all
these mechanisms may be implemented in a particular model in which
the custodial symmetry is imposed from the beginning. The radiative
correction mechanism may be implemented with scalar neutral bosons
or heavy leptons as well. In the latter case there are more than one
tadpole in the internal fermion line and the diagram is finite.

In fact, because the measured value of $\rho$ is, experimentally,
near 1, we know that the custodial symmetry of the SM is not so
badly broken. But, if this is an approximate symmetry of the full
model, its breakdown in the fermion and in the boson sectors may be
related to different sources (mechanisms) as we have shown in this
work. In all these cases the seesaw mechanism is mandatory in the
neutrino sector in order to generate small masses for the active
neutrinos.

\acknowledgments

This work was partially supported by CNPq under the processes
305185/03-9 (J.C.M.) and 306087/88-0 (V.P.).



\begin{thebibliography}{10}
\bibitem{volkas} R. R. Volkas, Phys. Rev. D \textbf{53}, 2681 (1996).
\bibitem{manohar} A. V. Manohar and C. T. Sachradja, Phys. Lett.
\textbf{B592}, 473 (2004).
\bibitem{veltman} M. Veltman, Nucl. Phys. \textbf{B123}, 89 (1977);
and Acta Phys. Pol. \textbf{B8}, 475 (1977).
\bibitem{custodial} P. Sikivie, L. Susskind, M. Voloshin, and V.
Zakharov, Nucl. Phys. \textbf{B173}, 189 (1980); R. S. Chivukula,
M. J. Dugan, M. Golden, and E. H. Simmons, Ann. Rev. Nucl. Part. Sci.
\textbf{45}, 255 (1993).
\bibitem{new} K. Agashe, R. Contino, L. Da Rold, and A. Pomarol,
Phys. Lett. \textbf{B641}, 62 (2006), hep-ph/0605341; M. Carena, E.
Pont\'{o}n, J. Santiago, and C. E. M. Wagner, hep-ph/0607106.
\bibitem{will} S. Willenbrock, Lectures presented at TASI 2004,
hep-ph/0410370 (unpublished).
\bibitem{seesaw} M. Gell-Mann, P. Ramond, and R. Slansky, in
\textit{Supergravity}, P. van Niewenhuizen and D. Z. Freedman, Eds.
(North-Holland, Amsterdam, 1979, pp. 315); T. Yanaguida, in
Proceedings of the \textit{Workshop on Unified Theory and Baryon
Number in the Universe}, O. Sawada and A. Sugamoto, Eds. (KEK,
Tsukuba, 1979); R. N. Mohapatra and G. Senjanovic, Phys. Rev. Lett.
\textbf{44}, 912 (1980).
\bibitem{sw72} S. Weinberg, Phys. Rev. Lett. \textbf{29}, 388 (1972).
\bibitem{newpprd} A. G. Dias, J. C. Montero, and V. Pleitez, Phys.
Rev. D \textbf{73}, 113004 (2006); hep-ph/0605051.
\bibitem{davidson} A. Davidson and K. Wali, Phys. Rev. Lett. \textbf{59},
393 (1987).
\bibitem{tau} J. C. Montero, C. A. de S. Pires, and V. Pleitez, Phys.
Rev. D \textbf{65}, 093017 (2002).
\bibitem{ll88} P. Langacker and D. London, Phys. Rev. D \textbf{38},
886, 907 (1988); E. Nardi, E. Roulet, and D. Tommasini, Phys. Lett.
\textbf{B344}, 225 (1995); T. C. Andre and J. L. Rosner, Phys. Rev.
D \textbf{69}, 035009 (2004); J. A. Aguilar-Saavedra, Phys. Lett.
\textbf{B625}, 234 (2005), Erratum-\textit{ibid}, \textbf{B633}, 762
(2006).
\bibitem{veltman77} M. Veltman, Nucl. Phys. \textbf{B123}, 89 (1977);
\textsl{Diagrammatica}, Cambridge University Press, Cambridge, 1994,
pp. 159--161.
\bibitem{nusright} J. C. Montero, F. Pisano, and V. Pleitez, Phys.
Rev. D \textbf{47}, 2918 (1993); H. Fanchiotti, C. Garcia-Canal, and
W. A. Ponce, Europhys. Lett. \textbf{72}, 733 (2005).
\bibitem{pp} F. Pisano and V. Pleitez, Phys. Rev. D \textbf{46},
410 (1992).
\bibitem{gw77} S. L. Glashow and S. Weinberg, Phys. Rev. D \textbf{15},
1958 (1977).

\bibitem{zprime} J. Hewett, and T. Rizzo, Phys. Rep. \textbf{183},
193 (1989); F. del Aguila, Acta Phys. Polo. \textbf{B25}, 1317
(1994); A. Leike, Phys. Rep. \textbf{317}, 143 (1999); X-G. He and
G. Valencia, Phys. Rev. D \textbf{68}, 033011 (2003).

\bibitem{georgiglashow} H. Georgi and S. L. Glashow, Phys. Rev. D
\textbf{7}, 2457 (1973).
\bibitem{valencia} X-G. He and G. Valencia, Phys. Rev. D \textbf{74},
013011 (2006).
\bibitem{cheung} K. Cheung, C-W. Chiang, N. G. Deshpande, and J.
Jiang, hep-ph/0604223.
\end{thebibliography}
\end{document}